\documentclass[8.5pt,twoside,twocolumn]{article}
\usepackage[super,sort&compress,comma]{natbib} 
\usepackage{mhchem}
\usepackage{times,mathptmx}
\usepackage{sectsty}
\usepackage{balance} 

\usepackage{graphicx} 
\usepackage{lastpage}
\usepackage[format=plain,justification=raggedright,singlelinecheck=false,font=small,labelfont=bf,labelsep=space]{caption} 
\usepackage{fancyhdr}
\usepackage[utf8]{inputenc}
\usepackage{hyperref}
\usepackage{graphicx}
\usepackage{amsmath,amssymb}

\usepackage{color}

\begin{document}

\thispagestyle{plain}
\fancypagestyle{plain}{
\fancyhead[L]{\includegraphics[height=8pt]{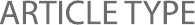}}
\fancyhead[C]{\hspace{-1cm}\includegraphics[height=20pt]{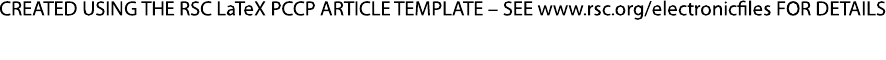}}
\fancyhead[R]{\includegraphics[height=10pt]{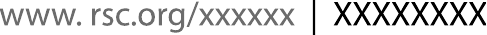}\vspace{-0.2cm}}
\renewcommand{\headrulewidth}{1pt}}
\renewcommand{\thefootnote}{\fnsymbol{footnote}}
\renewcommand\footnoterule{\vspace*{1pt}%
\hrule width 3.4in height 0.4pt \vspace*{5pt}} 
\setcounter{secnumdepth}{5}

\makeatletter 
\def\subsubsection{\@startsection{subsubsection}{3}{10pt}{-1.25ex plus -1ex minus -.1ex}{0ex plus 0ex}{\normalsize\bf}} 
\def\paragraph{\@startsection{paragraph}{4}{10pt}{-1.25ex plus -1ex minus -.1ex}{0ex plus 0ex}{\normalsize\textit}} 
\renewcommand\@biblabel[1]{#1}            
\renewcommand\@makefntext[1]%
{\noindent\makebox[0pt][r]{\@thefnmark\,}#1}
\makeatother 
\renewcommand{\figurename}{\small{Fig.}~}
\sectionfont{\large}
\subsectionfont{\normalsize} 

\fancyfoot{}
\fancyfoot[LO,RE]{\vspace{-7pt}\includegraphics[height=9pt]{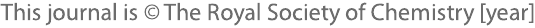}}
\fancyfoot[CO]{\vspace{-7.2pt}\hspace{12.2cm}\includegraphics{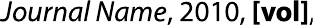}}
\fancyfoot[CE]{\vspace{-7.5pt}\hspace{-13.5cm}\includegraphics{RF}}
\fancyfoot[RO]{\footnotesize{\sffamily{1--\pageref{LastPage} ~\textbar  \hspace{2pt}\thepage}}}
\fancyfoot[LE]{\footnotesize{\sffamily{\thepage~\textbar\hspace{3.45cm} 1--\pageref{LastPage}}}}
\fancyhead{}
\renewcommand{\headrulewidth}{1pt} 
\renewcommand{\footrulewidth}{1pt}
\setlength{\arrayrulewidth}{1pt}
\setlength{\columnsep}{6.5mm}
\setlength\bibsep{1pt}

\twocolumn[
  \begin{@twocolumnfalse}
\noindent\LARGE{\textbf{Feedback control of inertial microfluidics using axial control forces}}
\vspace{0.6cm}

\noindent\large{\textbf{Christopher Prohm,$^{\ast}$ and Holger Stark}}\vspace{0.5cm}

\noindent\textit{\small{\textbf{Received Xth XXXXXXXXXX 20XX, Accepted Xth XXXXXXXXX 20XX\newline
First published on the web Xth XXXXXXXXXX 200X}}}

\noindent \textbf{\small{DOI: 10.1039/b000000x}}
\vspace{0.6cm}

\noindent \normalsize{
Inertial microfluidics is a promising tool for many lab-on-a-chip applications. Particles in channel flows with Reynolds 
numbers above one undergo cross-streamline migration to a discrete set of equilibrium positions
in square and rectangular channel cross sections.
This effect has been used extensively for particle sorting and the analysis of particle properties.
Using the lattice Boltzmann method, we determine equilibrium positions in square and rectangular cross sections
and classify their types of stability for different Reynolds numbers, particle sizes, and channel aspect ratios. 
Our findings thereby help to design microfluidic channels for particle sorting.
Furthermore, we demonstrate how an axial control force, which slows down the particles,
shifts the stable equilibrium position towards the channel center.
Ultimately, the particles then stay on the centerline for forces exceeding a threshold value. This effect is sensitive to
particle size and channel Reynolds number and therefore suggests an efficient method for particle separation.
In combination with a hysteretic feedback scheme, we can even increase particle throughput.
}
\vspace{0.5cm}
 \end{@twocolumnfalse}
  ]


\footnotetext{\textit{Institute of Theoretical Physics, Technische Universit\"at  Berlin, Hardenbergstr. 36, 10623 Berlin, Germany; 
	E-mail: Christopher.Prohm@TU-Berlin.de}}


\section{Introduction}

In recent years a number of devices using fluid inertia in microfluidic setups have been proposed for 
applications such as particle steering and sorting or for the whole range of flow cytometric tasks in biomedical
applications. 
They include cell counting, cell sorting, and mechanical phenotyping
\cite{Hur2011,Mach2010,Guan2013Spiral,Dudani2013Pinched}.  
These devices rely on cross-streamline migration of solute particles subject to fluid flow where fluid inertia 
cannot be neglected, as is commonly done in microfluidics.
In this article we demonstrate how control forces along the channel axis influence inertial cross-streamline migration
and how feedback control using axial forces enhances particle throughput.

Segr\'{e} and Silberberg, who investigated colloidal particles in circular channels, were the first to attribute
cross-streamline migration to fluid inertia \cite{Segre1961}. They observed that 
flowing particles gathered on a circular annulus about halfway between channel center and wall.
This effect is connected to an inertial lift force in radial direction. 
It becomes zero right on the annulus which marks degenerate stable equilibrium positions in the circular cross section. 
%
For microfluidic applications channels with a rectangular cross section are used since they can be fabricated more easily.
The reduced symmetry qualitatively changes the lift force profile
and only a discrete set of equilibrium positions remain \cite{DiCarlo2009}.
In square channels 
they are typically found halfway between the channel center and the centers of the channel walls \cite{DiCarlo2009Inertial}.
In numerical studies also migration to positions on the diagonals are observed \cite{Kataoka2011Numerical,Chun2006}.
In rectangular channels
the number of equilibrium positions is further reduced to two
when the aspect ratio strongly deviates from one \cite{Hur2011,Zhou2013Fundamentals}.
The particles all collect in front of the long channel walls.
The exact equilibrium positions are of special importance, as they ultimately determine 
how devices function based on inertial microfluidics \cite{Hur2011,Guan2013Spiral,Mach2010}.

Inertial lift forces that drive particles away from the channel center are caused by the non-zero curvature or
parabolic shape of the Poiseuille flow profile \cite{DiCarlo2009,Matas2004}.
Only close to the channel walls, wall-induced lift forces  push particle towards the center.
In channels with rectangular cross section the cuvature of the flow profile is strongly modified.
Along the short main axis the flow profile remains approximately parabolic, while along the long main axis it
almost assumes the shape of a plug flow with strongly reduced curvature in the center when 
the cross section is strongly elongated \cite{BookBruus}.
The large difference in cuvature along the two main axes modifies the lift force profiles in both 
directions\cite{DiCarlo2009}.
We will investigate them in more detail in this article.

The method of matched asymptotic expansion allows an analytic treatment of inertia-induced migration and
to calculate lift force profiles \cite{Asmolov1999,Ho1974}.
As the method requires the particle radius to be much smaller than the channel diameter,
it is hardly applicable to microfluidic particle flow, where this assumption is often violated.
Here, numerical approaches provide further insight.
Previous studies in three dimensions used the lattice Boltzmann method \cite{Chun2006}, the finite element method \cite{DiCarlo2009}, 
or multi-particle collision dynamics \cite{Prohm2012}. 

Using additional control methods such as optical lattices \cite{MacDonald2003} or optimal control
\cite{Prohm2013} can increase the efficiency of microfluidic devices.
In an attractive experiment Kim and Yoo demonstrate a method to focus particles to  the channel center
\cite{Kim2009}.
They apply an electric field along the channel axis to slow down the particles relative to 
the external Poiseuille flow, which induces a Saffmann force towards the channel center \cite{Saffman1965}.
The experiments are performed at Reynolds number $\mathrm{Re} \approx 0.05$. 
We will take up this idea, and study at moderate Reynolds numbers how the inertial lift force profile changes under an axial control force.

A more sophisticated method to operate a system is feeback control where the control action
depends on the current state. It is widely used in engineering and everyday life\ \cite{BookAstrom2010Feedback}.
In microfluidic systems optical tweezers combined with feedback control provide a strategy to
measure microscopic forces in polymers and molecular motors \cite{Jonas2008Light,Wuite2000Single,Wang1997Stretching}.
In lab-on-a-chip devices several strategies are suggested for sorting particles. 
They all monitor particle flow directly and use the recorded signal to implement feedback-controlled optical manipulation 
\cite{Applegate2007Optically,Wang2011Enhanced,Munson2010Image}.
We will apply a simple form of feedback control to keep particles in the channel center.

In this paper we use the lattice-Boltzmann method to investigate several aspects of inertial microfluidics. We 
study in detail the equilibrium particle positions in microfluidic channels with square and rectangular 
cross sections and categorize their types of stability. In particular, we show how for channels with sufficiently 
elongated cross sections, colloidal particles are constrained to move in a plane. We also show how the 
inertial lift force profile is manipulated by applying an axial control force such that the stable equilibrium
position gradually moves to the channel center. 
The effect strongly depends on particle size and therefore
can be applied for particle sorting. Finally, using the axial force we implement hysteretic feedback control
to keep the particle close to the channel center and demonstrate how this enhances particle throughput
compared to the case of constant forcing.

The article is organized as follows. In Sect.\ \ref{sec:methods} we introduce the microfluidic geometry, explain 
details of the lattice-Boltzmann implementation, and shortly introduce Langevin dynamics simulations. Our 
results on equilibrium positions and lift-force profiles in square and rectangular channels are reported in 
Sect.\ \ref{sec:uncontrolled}. We demonstrate the influence of axial control forces on the lift force profile in 
Sect.\ \ref{sec:control} and combine it with feedack control in Sec.\ \ref{sec:feedback}. We finish with 
conclusions in Sect.\ \ref{sec.concl}.

\section{Methods}
\label{sec:methods}

In this section we first introduce the microfluidic system we investigate in the following.
We then discuss the lattice Boltzmann method, the procedure used to determine the inertial lift forces, 
and finally the Langevin dynamics for our feedback-control scheme.

\subsection{Microfluidic system}
 
\begin{figure}
\includegraphics[width=\columnwidth]{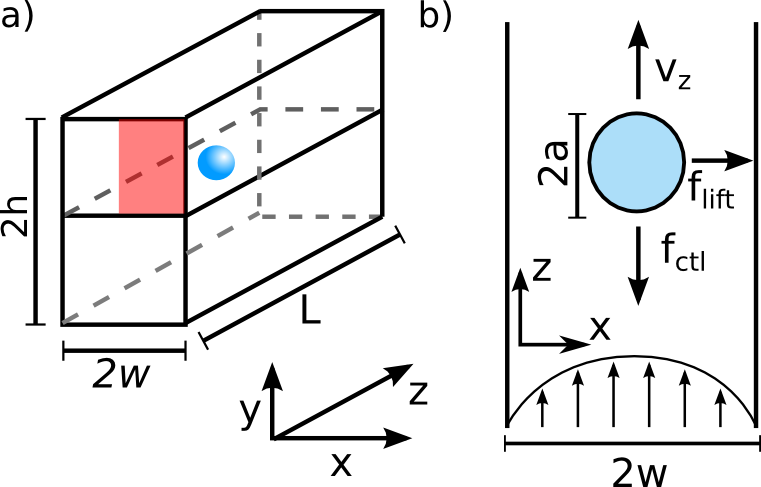}
\caption{A schematic of the microfluidic channel (a)
and the $xz$ plane at $y=0$ (b).
Further explanations are given in the main text.
It is sufficient to only determine the inertial lift force $f_\mathrm{lift}$ in the
red quadrant due to the symmetry of the rectangular cross section.
}
\label{fig:schematic}
\end{figure}

We investigate a microfluidic
channel with rectangular cross section of height $2h$, width $2w$, and length $L$ as illustrated 
in Fig.~\ref{fig:schematic}. We choose the coordinate system such that the $z$ axis coincides with the channel axis
and the $x$ and $y$ axis define the horizontal and vertical direction in the cross section, respectively.
The channel center corresponds to $x=y=0$.  
The channel is filled by a Newtonian fluid with density $\rho$ and kinematic viscosity $\nu$ 
and a pressure driven Poiseuille flow is applied \cite{BookBruus}. 
The maximum flow velocity $u_0$ at the channel center determines the Reynolds number 
$\mathrm{Re} = 2w u_0 / \nu$.
The implementation of the Poiseuille flow within the lattice Boltzmann method will be discussed in the next section.

Inside the channel we place a neutrally buoyant colloid with radius $a$.
It follows the streamlines of the applied Poiseuille
flow with an axial velocity $v_z$ close the external Poiseuille flow velocity.
Due to the fluid inertia the colloidal particle experiences a lateral lift force $f_\mathrm{lift}$,
which leads to cross-streamline migration.
In Sects.~\ref{sec:control} and \ref{sec:feedback} we also apply
an additional axial control force $f_\mathrm{ctl}$ to the colloidal particle.
We use periodic boundary conditions along the axial direction and a channel length of $L = 20 a$ 
to ensure that the periodic colloidal images do not interact with each other and thereby do not
influence our results.

\subsection{The lattice Boltzmann method}

We use the lattice Boltzmann method (LBM) to solve the Navier-Stokes equations of a
Newtonian fluid \cite{Duenweg2008Lattice,Aidun2010Lattice}.
LBM employs an ensemble of point particles that perform alternating
steps of free streaming and collisions.
The particles are constrained to move on a cubic lattice with lattice spacing $\Delta x$.
This restricts the particle velocities to a discrete set of vectors $\vec{c}_i$ such that after each
streaming step with duration $\Delta t$ the new particle positions again lie on the lattice.
In LBM one describes the number of particles at lattice point $\vec{x}$ with velocity $\vec{c}_i$ by the 
distribution function $f_i(\vec{x}, t)$.
The first two moments of this distribution function give the hydrodynamic variables: number density $\rho(\vec{x}, t)
 = \sum_i f_i(\vec{x}, t)$
and velocity $\vec{u} (\vec{x}, t)= \frac{1}{\rho} \sum_i \vec{c}_i f_i(\vec{x}, t)$.

In the collision step the fluid particles at each lattice point
exchange momentum by a local collision rule.
Here we employ the common Bhatnagar-Gross-Krook collision model, where the velocity
distribution function relaxes towards a local equilibrium distribution $f_i^\mathrm{eq}$ with a single 
relaxation time $\tau$. 
This results in the post-collision distribution
\begin{align}
	f_i^\star(\vec{x}, t) = f_i(\vec{x}) + \frac{1}{\tau} \left[ f_i^\mathrm{eq}(\vec{x}, t) - f_i(\vec{x}, t) \right].
\end{align}
For the local thermal equilibrium distribution we use an expansion of the local Maxwell-Boltzmann distribution 
up to second order in the mean velocity $\vec{u}$, which results in
\begin{align}
	f_i^\mathrm{eq} &= w_i \rho \left(
			1 + 
			\frac{\vec{c}_i \cdot \vec{u}}{c_s^2} + 
			\frac{(\vec{c}_i \cdot \vec{u})^2}{2 c_s^4} -
			\frac{|\vec{u}|^2}{2 c_s^2} 
		\right).
\end{align}
The weights $w_i$ ensure that all moments of the equilibrium distribution up to the third order are correctly
reproduced including the number density $\rho$ (zeroth order) and the mean velocity $\vec{u}$ at lattice 
point $\vec{x}$ (first order).
$c_s = \sqrt{k_B T/m}$ is the speed of sound \cite{Duenweg2008Lattice}.
Note that the collision step locally conserves mass and momentum.

After collision the fluid particles move to adjacent lattice positions 
according to their velocities and the new distribution functions at time $ t + \Delta t$ become
\begin{align}
	f_i(\vec{x} + \Delta t \vec{c}_i, t + \Delta t) = f^\star_i(\vec{x}, t).
\end{align}
Here we use the D3Q19 scheme \cite{Duenweg2008Lattice}, where the velocities connect each lattice point to its nearest and next-nearest neighbors. 
To simplify the following discussion, in the remainder of this section we set $\Delta x = 1$ and $\Delta t = 1$
and rescale all quantities accordingly.

Note that on length and time scales much larger than $\Delta x$ and $\Delta t$, respectively,
one can derive the Navier-Stokes equation by a Chapman-Enskog expansion using the formulated streaming and collision steps
\cite{Duenweg2008Lattice}. 
The internal pressure follows an ideal gas law with $p = c_s^2 \rho$, where $c_s = 1 / \sqrt{3}$ 
is the speed of sound, and the kinematic viscosity is given by $\nu = c_s^2 (\tau - 1 / 2)$.

To implement the no-slip boundary condition on the channel walls, we employ the regularized
boundary condition introduced by Latt and Chopard \cite{Latt2008}.
It treats boundary nodes just like fluid nodes but modifies the distribution function 
before the collision such that the correct velocity is imposed. 
The method uses the bounce-back rule for the nonequilibrium distribution
introduced by Zou and He \cite{Zou1997Pressure}.

\begin{figure}
\includegraphics[width=\columnwidth]{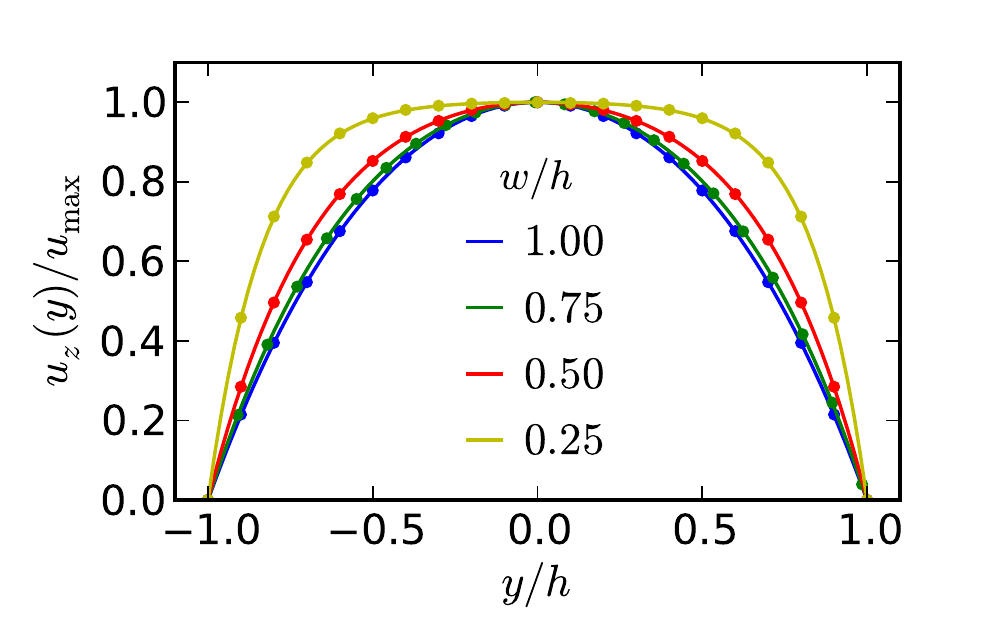}
\caption{Velocity profiles of the Poiseuille flow in the rectangular channel
		plotted at $x=0$ along the $y$ axis (the long cross sectional axis) for different aspect ratios $w/h$.
		 The solid lines show the analytical form of the profiles \cite{BookBruus}, while the symbols show
		 results of the LB simulations.
}
\label{fig:poiseuille-flow}
\end{figure} 

We implement the pressure driven Poiseuille flow by imposing a constant body force $\vec{g}$ on 
the fluid such that the fluid velocity $\vec{u}(\vec{x},t)$ used to calculate the equilibrium distribution $f_i^\mathrm{eq}$
is replaced by \cite{Shan1993}
\begin{align}
	\vec{u} \rightarrow \vec{u} + \tau \vec{g}.
\end{align}
We confirm in Fig.~\ref{fig:poiseuille-flow} that this procedure does indeed reproduce the 
analytically known Poiseuille flow profile.

We place a colloid in the Poiseuille flow and study its position $\vec{r}$, velocity $\vec{v}$, and angular velocity $\vec{\omega}$.
We couple the colloid to the fluid using the Inamuro Immersed Boundary (IB) method \cite{Inamuro2012Lattice} with ``five iterations''.
For reference we present a short summary of our implementation.

The colloid surface is approximated by a triangular mesh with vertices $i$ at positions  $\vec{x}^m_i$. 
To obtain the mesh, we start from a an icosahedron and successively refine it
by splitting each triangle into four until the edge length is smaller than the lattice spacing. 
The positions of the resulting vertices continuously vary
in space and hence do not necessarily coincide with the lattice sites. 
For clarity we will denote here the lattice sites by $\vec{x}_j$.
We couple mesh vertices and lattice sites to each other using a smoothed delta function $\delta_h(\vec{x})$.
We follow Peskin \cite{Peskin2002} and employ
$ \delta_h(\vec{x}) = \phi(x)\phi(x)\phi(z) $ with 
\begin{align}
	\phi(x) &= 
		\begin{cases}
			\frac{1}{8} \left(3 - 2 |x| + \sqrt{1 + 4|x| - 4x^2}\right) & 0 \leq |x| \leq 1 \\
			\frac{1}{8} \left(5 - 2 |x| - \sqrt{7 + 12|x| - 4x^2}\right) & 1 \leq |x| \leq 2 \\
			0 & 2 \leq |x|
		\end{cases}
\end{align}
and the same form for $\phi(y)$ and $\phi(z)$.
In the IB method one determines the fluid velocity $\vec{u}^{m}_i$ at mesh point $i$ by 
interpolating the fluid velocity from the lattice sites with the help of the smoothed delta function,
\begin{align}
	\vec{u}^{m}_i &= \sum_j \delta_h(\vec{x}^m_i - \vec{x}_j) \vec{u}_j.
\end{align} 
%
To enforce the no-slip boundary condition at the colloid surface,
we introduce the penalty force $\vec{f}^m_i = \vec{u}^m_i - \vec{v}^{s}_i$ 
as the difference between the fluid velocity and the surface velocity $\vec{v}^{s}_i = \vec{v} + \vec{\omega} \times ( \vec{x}^m_i - \vec{r})$ 
of the colloid at the position $\vec{x}^m_i$ of mesh vertex $i$.
The penalty force $\vec{f}^m_i$ acts on the mesh vertex $i$ and, 
to conserve momentum, its negative $-\vec{f}^m_i$ acts on the surrounding fluid.
We interpolate the penalty force on lattice site $\vec{x}_j$ from the neighboring mesh vertices,
\begin{align}
\vec{f}_j &= - \sum_i \delta_h(\vec{x}^m_i - \vec{x}_j) \vec{f}^m_i.     
\end{align}
To apply the penalty force to the fluid, we use the same method as for the body force. 
We calculate modified fluid velocities at the mesh points $i$, which do not obey the no-slip boundary 
condition exactly since we interpolate forces and velocities between the mesh and lattice points.
To decrease the slip velocity further, we therefore refine the penalty force iteratively by repeating the 
procedure five times and thereby implement the no-slip boundary condition in good approximation.
Note that the total penalty force experienced by the fluid and hence by the colloid is the sum over all iterations.

As just introduced, the fluid interacts with a colloid which results in a hydrodynamic coupling.
We can quantify it by a force and torque acting on the colloid given by the sum of the vertex contributions
just introduced,
\begin{align}
	\vec{F}_\mathrm{fluid} &= \sum_i \vec{f}^m_i    \label{eqn:fluid-force}  \\
	\vec{T}_\mathrm{fluid} &= \sum_i \left( \vec{x}^m_i - \vec{r} \right) \times \vec{f}^m_i  .
\end{align} 
The force and torque contain two contributions. The first one comes from the fluid particles outside the colloid.
The second contribution resulting from fluid particles inside the colloid is unphysical. We therefore  compensate this contribution 
using Feng's rigid body approximation \cite{Feng2009Robust} and denote the respective force and torque by 
$\vec{F}_\mathrm{Feng}$, $\vec{T}_\mathrm{Feng}$.

With all force contributions the equations of motion for the colloid are given by 
\begin{align}
	  \vec{r}(t + 1) &= \vec{r}(t) + \vec{v}(t),  \nonumber\\
 	M \vec{v}(t + 1) &= M \vec{v}(t) + \vec{F}_\mathrm{fluid} + \vec{F}_\mathrm{Feng} + \vec{F}_\mathrm{ctl}, 
	\label{eq.colloid}	\\
 	I \vec{\omega}(t + 1) &= I \vec{\omega}(t) + \vec{T}_\mathrm{fluid} + \vec{T}_\mathrm{Feng},   \nonumber
\end{align} 
where $M$ and $I$ 
are the respective mass and moment of inertia of the colloid and $\vec{F}_\mathrm{ctl}$ is the axial control force which we will introduce in
Sects.~\ref{sec:control} and \ref{sec:feedback}.

We use the palabos LB code \cite{PalabosWebsite} to implement the LB algorithm. 
We modified the immersed boundary (IB) algorithm to correctly account for periodic 
boundary conditions along the channel axis
and implemented the colloid dynamics of Eqs.\ (\ref{eq.colloid}).

Along the channel width we use a total of 101 lattice sites including the boundaries.
We implement  a cubic simulation grid and choose the number of lattice sites in the other
two directions accordingly.
We choose the maximum flow velocity in the channel such that the Mach number satisfies $\mathrm{Ma} = u_\mathrm{max} / c_s \leq 0.1$.
Finally, we adjust the kinematic viscosity by the the relaxation time $\tau$
and thereby fix the
desired Reynolds number $\mathrm{Re} = u_\mathrm{max} w / \nu$.
When $\tau > 1$, we readjust the Mach number such that $\tau = 1$, as it has been shown that
the accuracy of the combined LBM-IB methods greatly degrades 
for relaxation times larger than one  \cite{Le2009Boundary,Kruger2009Shear}.


\subsection{Determing inertial lift forces}

To determine inertial lift forces from LB simulations, we constrain the colloid to a fixed lateral position
by simply disregarding any colloid motion in the cross-sectional plane. 
However, we do do not impose any constraints on the axial and rotational motion and then determine the steady state in the LB simulations.

To speed up our simulations,
we initialize the system with the analytical solution of the rectangular Poiseuille flow and give the 
colloid an initial axial velocity of $v_z = 0.8 u_0$, where $u_0$ is the flow velocity at the channel center. 
Going through a transient dynamics, the system relaxes rapidly into a unique steady state
within the first 1000 time steps. We continue the time-evolution up to 
the vortex diffusion time $T = 0.5 w^2 / \nu$ and determine the inertial lift force by averaging the colloidal 
force $\vec{F}_\mathrm{fluid}$ from Eq.\ (\ref{eqn:fluid-force}) over the last 2000 time steps of the simulation.
We demonstrated before \cite{Prohm2013,Prohm2012} that this procedure does indeed reproduce correct lift-force profiles.

\subsection{Langevin dynamics simulations}
\label{sec:langevin-equation}

As demonstrated below, axial control forces influence the inertial lift-force profiles which we
determine in LB simulations. We then use these  profiles in Langevin dynamics simulations of the colloidal
motion to investigate the potential benefit of feedback control using axial control forces.

We will restrict ourselves to channels with an aspect ration $w/h = 1/3$, which ensures that the
colloidal dynamics essentially takes place in the $xz$ plane as discussed in Sect.\ \ref{sec:rectangular}.
As we will show in Sect.\ \ref{sec:control}, the inertial lift force $f_\mathrm{lift}$ and the 
axial velocity $v_z$ depend on the applied axial control force $f_\mathrm{ctl}$.
We also include thermal noise to exploit the stability of the fix points of colloidal motion under feedback control.
Following our work in \cite{Prohm2013}, we only include thermal noise along the lateral direction, as 
the axial velocities are much larger than the lateral ones.
Then, the Langevin equations of motion in lateral and axial directions are given by 
\begin{align}
	\xi \frac{\mathrm{d}}{\mathrm{d} t} x &= f_\mathrm{lift}(x, f_\mathrm{ctl}) + \eta(t), \\
	 	\frac{\mathrm{d}}{\mathrm{d} t} z &= v_z(x, f_\mathrm{ctl}),
\end{align}
where the white noise force has zero mean, $<\eta(t)> = 0$, and its variance obeys the fluctuation-dissipation theorem,
$<\eta(t) \eta(t^\prime)> = 2 k_B T \xi \delta(t - t^\prime)$.
We solve the Langevin equations using the conventional Euler scheme \cite{BookKloeden}. 
The parameters are chosen for a channel with width $2w = 20\mathrm{\mu m}$ and temperature $T = 300K$.

\section{Inertial lift forces for different channel geometries}
\label{sec:uncontrolled}

\begin{figure}
\includegraphics[width=\columnwidth]{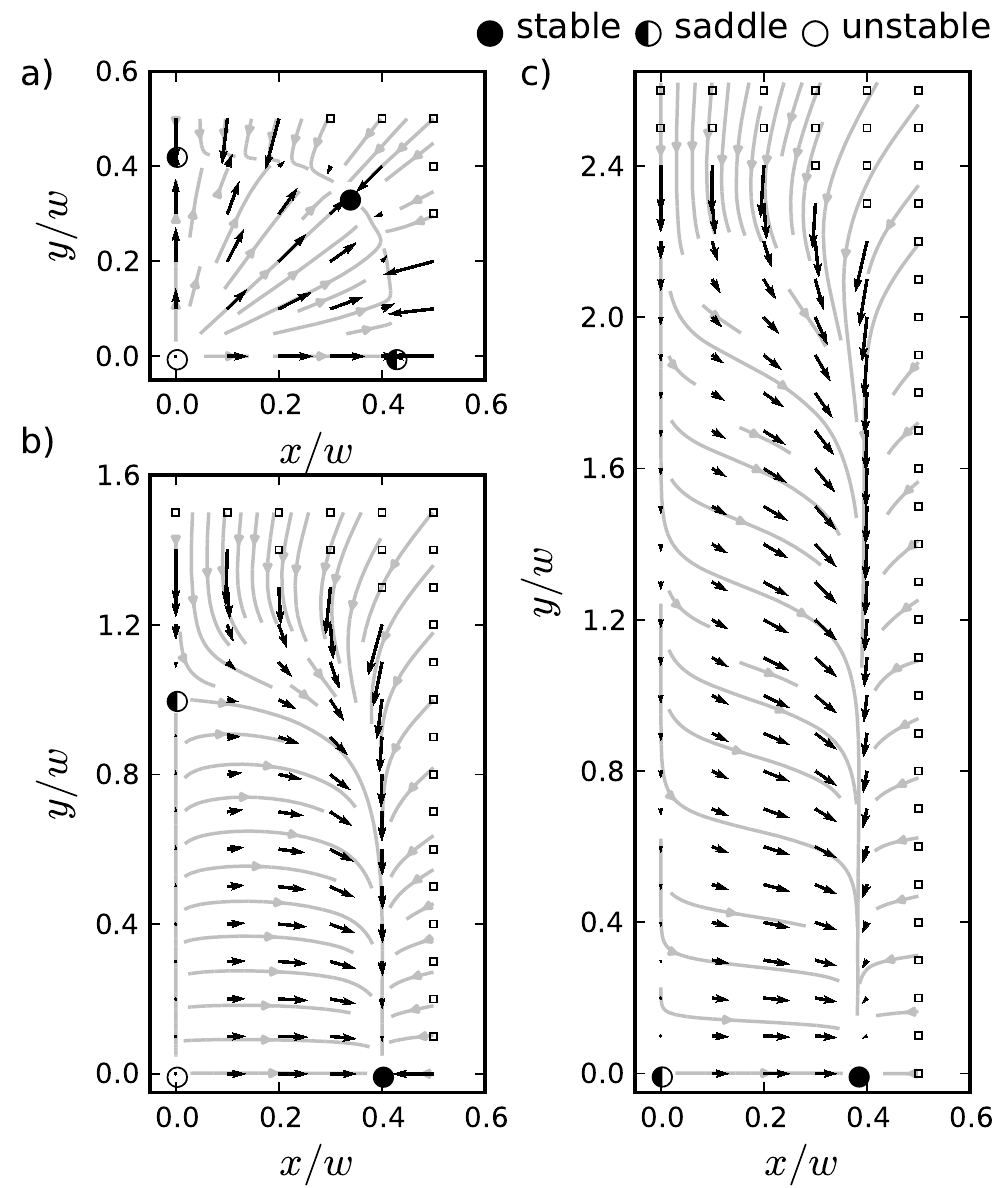}
\caption{Inertial lift forces (black arrows) for a colloidal particle with radius $a/w = 0.4$ 
		in a pressure driven flow at $\mathrm{Re} = 10$.
		The forces are plotted in the upper right quadrant of the microchannel cross section
		for aspect ratios $h/w = 1$ (a), $h/w = 2$ (b) and $h/w = 3$ (c).
		The gray lines indicate the trajectories of the colloidal particle as it experiences the lift forces. 
		Lift forces larger than $0.35\rho\nu^{2}$ are not shown, their positions are indicated by squares.
		Also indicated are stable and unstable equilibrium positions. With ``saddle'' we denote equilibrium
		positions only unstable along one direction.
		  }
\label{fig:cross-sections}
\end{figure}


In channels with circular cross sections, inertial lift forces drive colloids to a
circular annulus with a radius of about half the channel radius. 
The axial symmetry is reduced in channels with square or rectangular cross sections and instead of an annulus particles 
accumulate at a discrete set of stable equilibrium positions \cite{DiCarlo2009Inertial}. 
In addition, the system also shows unstable equilibrium positions, where the lift force also vanishes but particles migrate away from 
them upon a small disturbance.

In the following two sections we investigate the location and the stability of the equilibrium positions for 
different particle sizes, Reynolds numbers, and channel geometries.
We discuss in detail how we can tailor colloidal motion by varying the aspect ratio of the channel cross section.
Due to symmetry, we can restrict our discussion to the upper right quadrant shown in Fig.\ \ref{fig:schematic}.
In Fig.~\ref{fig:cross-sections} we show the forces acting on a particle with radius $a / w = 0.4$ and the resulting trajectories
at Reynolds number $\mathrm{Re} = 10$ for different channel cross sections.
We discuss the relevant features first for a channel with square cross section and then for 
general rectangular cross sections. 


\subsection{Square channels}
\label{sec:square}

In a channel with square cross section and at Reynolds number $\mathrm{Re}= 10$,
a particle with radius $a/w = 0.4$ experiences the inertial force profile shown in Fig.~\ref{fig:cross-sections}(a). 
The gray lines indicate possible trajectories followed by particles, which are free to migrate. 
Stable and unstable equilibrium positions are also indicated.
We observe that the migration roughly occurs in two steps.
From the channel center and the channel walls strong radial forces drive the particle onto an almost circular 
annulus at about $r \approx 0.4w$. 
Since the forces are strong, the migration occurs very rapidly.
Then the particle slowly migrates along the annulus to its equilibrium position, here situated on the diagonal direction.

\begin{figure}
\includegraphics[width=\columnwidth]{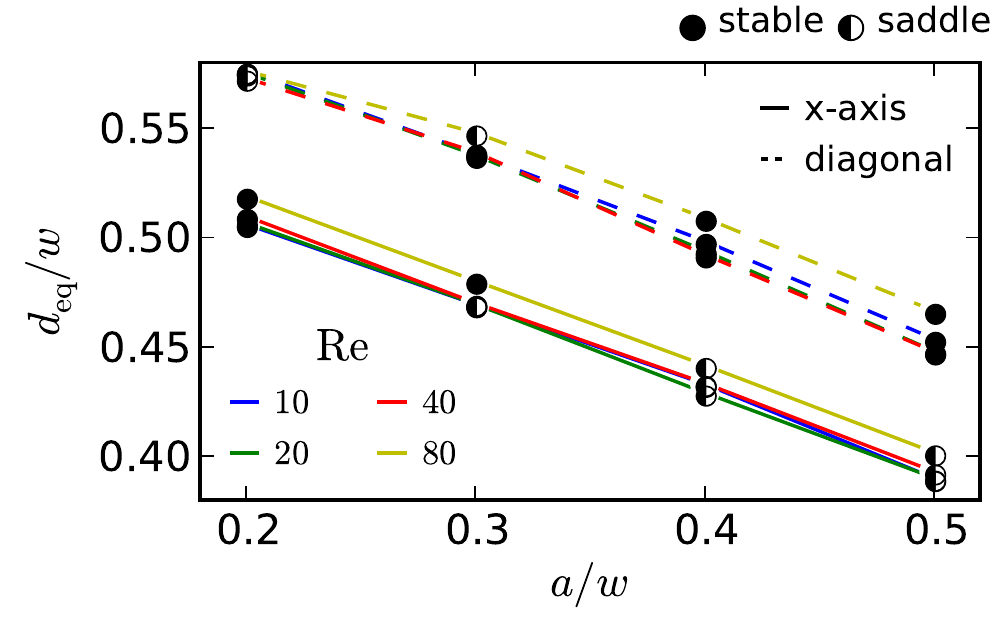}
\caption{
Equilibrium positions $d_{\mathrm{eq}}$ (distance from the center) in a square channel
plotted versus colloid radius for different Re. 
Equilibrium positions exist along the main axis in $x,y$ directions (solid lines) and along the diagonal (dashed line).
Closed circles indicate stable equilibrium positions, whereas open circles are unstable.
}
\label{fig:square-eq}
\end{figure}

Together with the channel center there are in total nine equilibrium positions or fix points in the channel
cross section. Four of them are indicated in Fig.\ \ref{fig:square-eq}(a). 
The channel center is always unstable and particle migrate away from it.
There are four fix points along the diagonal axes and four along the main axes ($x,y$ directions) of the channel cross section.
We plot their distances from the center versus colloid  radius for several Re in Fig.~\ref{fig:square-eq} and also indicate their stability.
The fix points along the diagonals are always positioned further away from the channel center as there is more space for the particle.
Consistent with previous results \cite{Chun2006,DiCarlo2009,Prohm2012}, we observe how both types of equilibrium 
positions move closer towards the channel center with increasing particle size and decreasing Reynolds number. 
Most importantly, small particles at high Reynolds numbers have their stable equilibrium positions
on the main axes, while larger particles at lower Reynolds number move  to the equilibrium positions on the diagonals.
This is a new result compared to previous treatments \cite{Chun2006,DiCarlo2009,Kataoka2011Numerical}.
   
In literature equilibrium positions in square channels have been reported along the main axes \cite{DiCarlo2009}, 
along the diagonals for large deformable drops \cite{Kataoka2011Numerical} or on both axes \cite{Chun2006}.
In contrast to \cite{Chun2006} we observe that particles move either to the diagonal equilibrium positions or to fixpoints
on the main axes but the equilibrium positions are never stable at the same time as illustrated in Fig.\ \ref{fig:square-eq}.
It has been demonstrated in spiral channels with trapezoidal cross section \cite{Guan2013Spiral} 
that such a sudden change in stability can be used to efficiently sort particles by size.
We note that while the particle size is fixed by the specific system under investigation,
the Reynolds number remains a free parameter and can be used to tune the stability of the equilibrium positions.


\subsection{Rectangular channels}
\label{sec:rectangular}

In experiments typically channels with rectangular cross sections are used
since the number of stable equilibrium positions reduces to two situated on the short main axis \cite{DiCarlo2009}. 
We observe the same behavior in the force profiles in Fig.~\ref{fig:cross-sections} for large colloids. 
While for channels with square cross section a particle migrates to its stable position on 
the diagonal  [Fig.~\ref{fig:cross-sections}(a)], 
this fixpoint vanishes with decreasing aspect ratio $w/h$ and the stable equilibrium position
switches to the short main axis along the $x$ direction [Fig.~\ref{fig:cross-sections}(b)]. 
Further decreasing the aspect ratio $w/h$, the saddle fixpoint on the $y$ axis vanishes completely
and moves to the center at $x=y=0$, where it keeps its stability along the $y$ axis  [Fig.~\ref{fig:cross-sections}(c)]. 
This has the important consequence (already exploited by us \cite{Prohm2013}) 
that the colloid is constrained to the 
$xz$ plane at $y=0$ and its dynamics becomes two-dimensional.
In contrast, in the situation of Fig.~\ref{fig:cross-sections}(b) a particle starting close to the centerline
moves out of the  $y=0$ plane on its way to the stable equilibrium position at $x \approx 0.4$. We now elaborate
in more detail on these observations.

\begin{figure}
\includegraphics[width=\columnwidth]{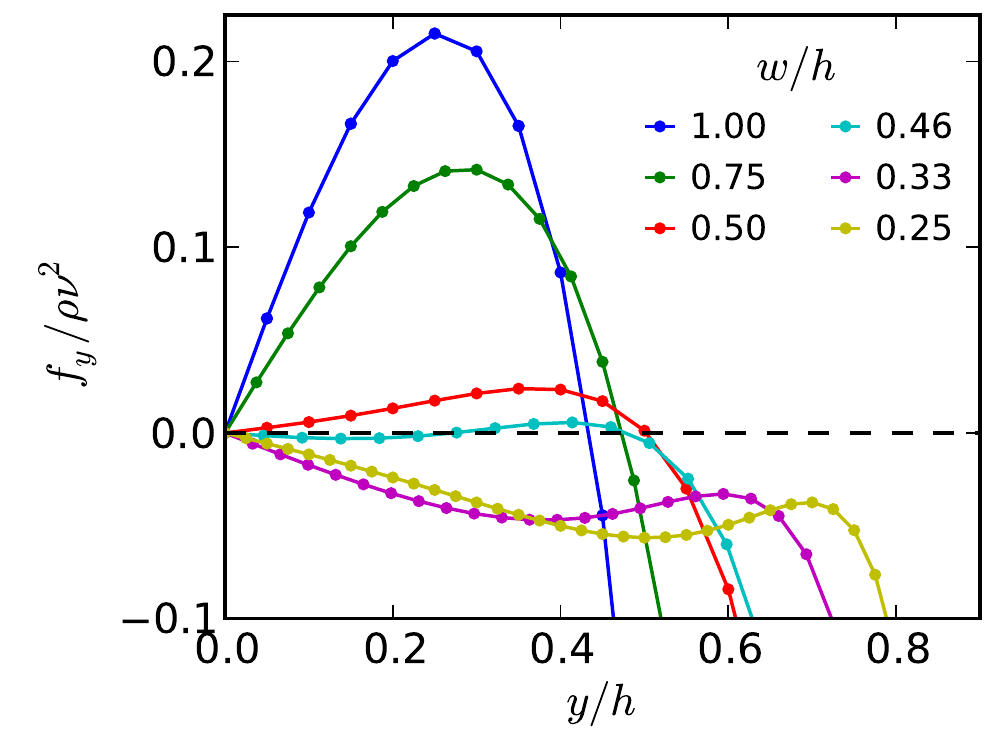}
\caption{%
Lift force $f_y$ along the $y$-direction  at $x=0$
plotted versus $y$ position for $\mathrm{Re} = 10$ and $a/w = 0.4$.
The different colors correspond to different aspect ratios $w/h$.}
\label{fig:force-y}
\end{figure} 

We first plot the lift force along the $y$ axis, \emph{i.e.}, at $x=0$ for several aspect ratios $w/h$ in Fig.~\ref{fig:force-y}. 
Due to symmetry, the lift force always points along the $y$ direction.
For the quadratic cross section, $w/h=1$, zero lift forces indicate the unstable fixpoint in the center ($x=y=0$)
and the saddle fixpoint at $y \approx 0.42$, which is unstable in $x$ direction. 
As the channel cross section elongates along the $y$ direction with decreasing $w/h$, the lift force driving the particle away
from the channel center becomes weaker and the saddle fixpoint shifts towards the channel wall.
Below a width $w \approx 0.45 h$, the lift force close to the center becomes negative and the unstable fixpoint
at $y=0$ splits into a saddle fixpoint (now stable in $y$ direction) and an additional unstable fixpoint. 
While we do not show this situation in Fig.\ \ref{fig:cross-sections}, it qualitatively looks the same as in the complete force profile in 
Fig.\ \ref{fig:control-tableau}(a) for smaller colloids. 
This is the onset, where the channel center becomes stable against motion along the $y$ axis and the colloid is constrained to the $xz$ plane at $y=0$. 
Further decreasing the aspect ratio $w/h$, the unstable and saddle fixpoints at $y \ne 0$ merge and vanish completely. 
Only the saddle fixpoint at $x=y=0$ remains as illustrated in Fig.~\ref{fig:cross-sections}(c).
Finally, we note that at $w/h \approx 0.75$ the stable equilibrium position in the channel cross section
switches from the diagonal to the $x$ axis, which is not observable in Fig.\ \ref{fig:force-y}.

\begin{figure}
\includegraphics[width=\columnwidth]{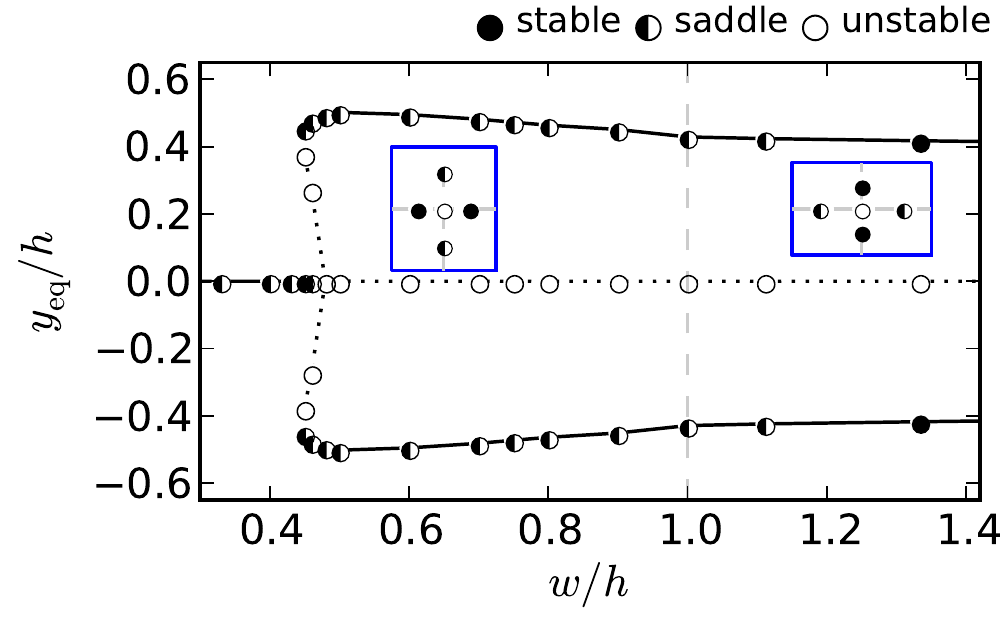}
\caption{Equilibrium positions along the $y$ axis plotted versus channel aspect ratio $w/h$
for $\mathrm{Re} = 10$ and colloid radius $a/w = 0.4$. The equilibrium positions are categorized
as stable, saddle, and unstable using also their stability with respect to the $x$ direction.
The blue insets show typical equilibrium positions in the channel cross section
for $w / h < 1$ (left) and for $w / h > 1.33$ (right).
}
\label{fig:bifurcation-diagram}
\end{figure}

\begin{figure}
\includegraphics[width=\columnwidth]{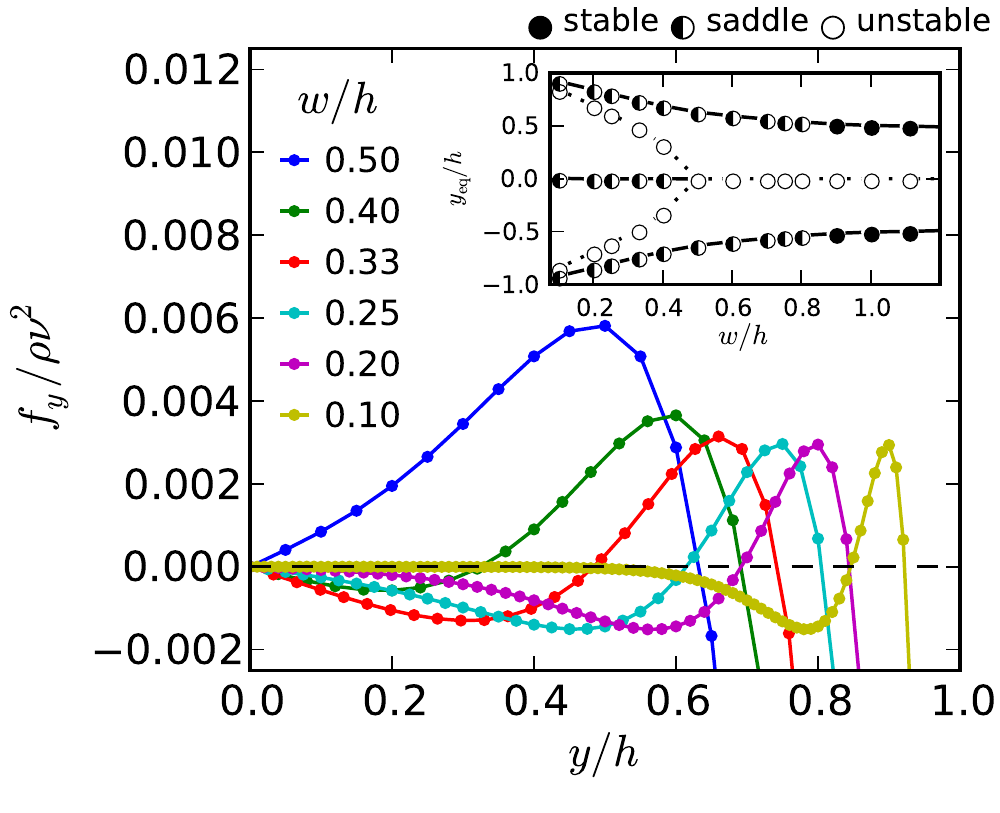}
\caption{
Lift force $f_y$ along the $y$ direction  at $x=0$
plotted versus $y$ position for $\mathrm{Re} = 10$ and $a/w = 0.2$.
The different colors correspond to different aspect ratios $w/h$.
The inset shows the equilibrium positions along the $y$ axis plotted versus channel aspect ratio $w/h$.
The equilibrium positions are categorized as stable, saddle, and unstable using also their stability with respect to the $x$ direction.
}
\label{fig:bifurcation-small}
\end{figure}

We summarize the situation in the bifurcation diagram of Fig.\ \ref{fig:bifurcation-diagram}, where we plot
the equilibrium positions on the $y$ axis versus the aspect ratio $w/h$. 
At sufficiently small $w/h$ only the saddle fixpoint at $y=0$ exists [Fig.~\ref{fig:cross-sections}(c)]. 
With increasing $w/h$ a subcritical pitchfork bifurcation occurs. 
A second fixpoint appears which splits into the saddle and unstable fixpoint [Fig.\ \ref{fig:control-tableau}(a)].
The latter ultimately merges with the fixpoint at $y=0$ which becomes unstable [Fig.~\ref{fig:cross-sections}(b)].
This resulting situation is illustrated in the left inset for the whole cross section and with the stable fixpoint on the $x$ axis. 
The stable fixpoint moves to the diagonal at $w/h \ge 0.75$.
The regime of the subcritical pitchfork bifurcation is much more pronounced for smaller particles as
illustrated in the inset of Fig.\ \ref{fig:bifurcation-small} for a colloid radius $a/w=0.2$. 
The lift force profiles in this regime (see Fig.\ \ref{fig:bifurcation-small})
even show that the subcritical transition to a single saddle fixpoint does not occur for small aspect ratios. This might be due to
the fact that below $w/h = 0.4$ the flow profiles close to the wall are the same.
Coming back to Fig.\ \ref{fig:bifurcation-diagram}. 
At $w/h > 1$ (now the short main axis points along the $y$ direction), the saddle fixpoint at nonzero $y_{\mathrm{eq}}$ first remains. 
It becomes stable at an aspect ratio $w/h \approx 1.33$, when the fixpoint on the diagonal vanishes. 
The situation is sketched in the right inset.

%
The basic features of the inertial lift force can be explained by considering the unperturbed flow field.
In particular, the lift force depends on the curvature of the flow field \cite{Matas2004,DiCarlo2009}.
Along the long channel axis ($y$ direction) we observe the flow velocity shown in Fig.~\ref{fig:poiseuille-flow}.
As the channel hight $h$ increases, the flow profile in the center flattens considerably and the curvature strongly increases.
The differences in the flow profiles for $w/h =1$ and $0.75$ are small, which corresponds to the modest decrease in the
strength of the lift force in Fig.\ \ref{fig:force-y}. At $w/h=0.5$ the pronounced flattening of the flow profile sets in which marks
the occurence of the subcritical bifurcation and the strong changes in the lift force profiles in Fig.\ \ref{fig:force-y}.
Finally, we note that the strength of the lift force in $x$-direction also becomes weaker with decreasing
aspect ratio $w/h$ but the overall characteristics of the profile (two fixpoints) remain the same.
Especially for aspect ratios $w/h \leq 0.5$, the lift force in $x$-direction does hardly change.

\begin{figure}
\includegraphics[width=\columnwidth]{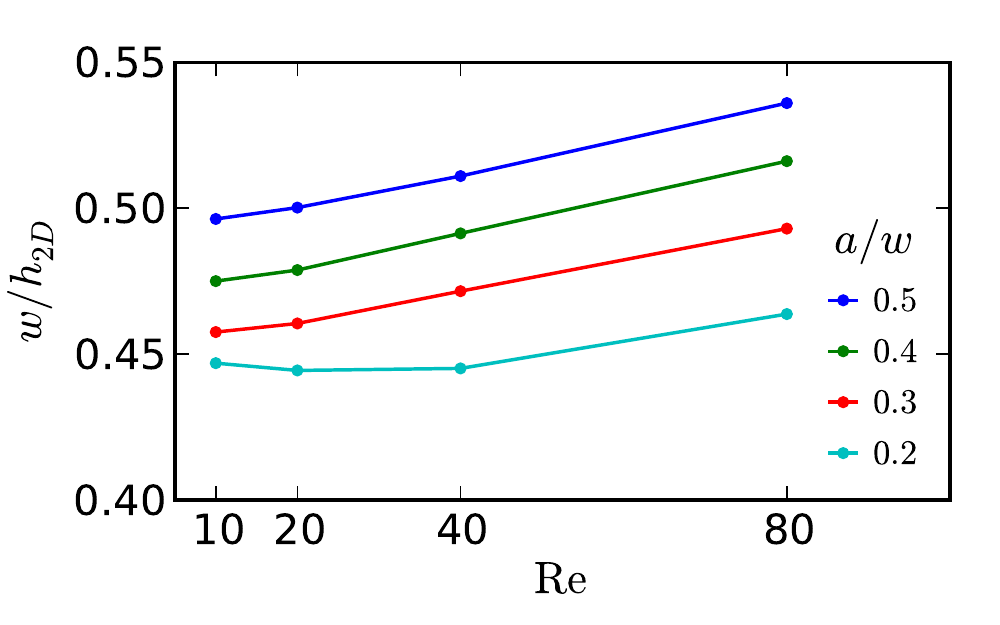}
\caption{
Aspect ratio $w/h_{\mathrm{2D}}$ at which the center becomes a saddle fixpoint and the
particles move in the center plane. $w/h_{\mathrm{2D}}$ is plotted versus Re for different 
particle sizes. The symbols are data points from the simulations.
}
\label{fig:necessary-aspect-ratio}
\end{figure}

For many microfluidic applications, such as cytometry \cite{Hur2010} or particle separation \cite{Mach2010}, 
it is advantageous to ensure that particles are constrained to move in a plane so that they can easily be monitored
in the focal plane of a microscope.
Here we observe that as soon as the center position becomes a saddle point, particles will not leave the center plane 
any more since they always experience a force driving them back towards the plane.
In Fig.\ \ref{fig:necessary-aspect-ratio} we plot the necessary aspect ratio to constrain particles to the center plane.
It becomes smaller with with decreasing particle size and Reynolds number.  
For the particle sizes investigated here it is sufficient to choose $w/h < 0.4$ to 
ensure that the system is effectively two-dimensional.

\section{Axial control of lift forces}
\label{sec:control}

In their experiments Kim and Yoo apply an axial electric field which slows down particles relativ to the 
Poiseuille flow \cite{Kim2009}. As a result,  particles are pushed towards the center line.
The observed migration can be rationalized with the Saffman force which is an inherent inertial force 
\cite{Saffman1965}. It acts perpendicular to a shear flow when particles are slowed down or sped up
relative to the fluid flow. In their experiment Kim and Yoo considered flow with channel Reynolds numbers 
$\mathrm{Re} \approx 0.05$ well below unity. Our idea is to apply this concept to moderate Re and manipulate
the inertial lift force using the additional Saffman force. We show that with the help of an axial control force, we 
can modify the inertial lift force profile such that we can steer a particle to almost any desired position on the 
$x$ axis.

\begin{figure}
\includegraphics[width=\columnwidth]{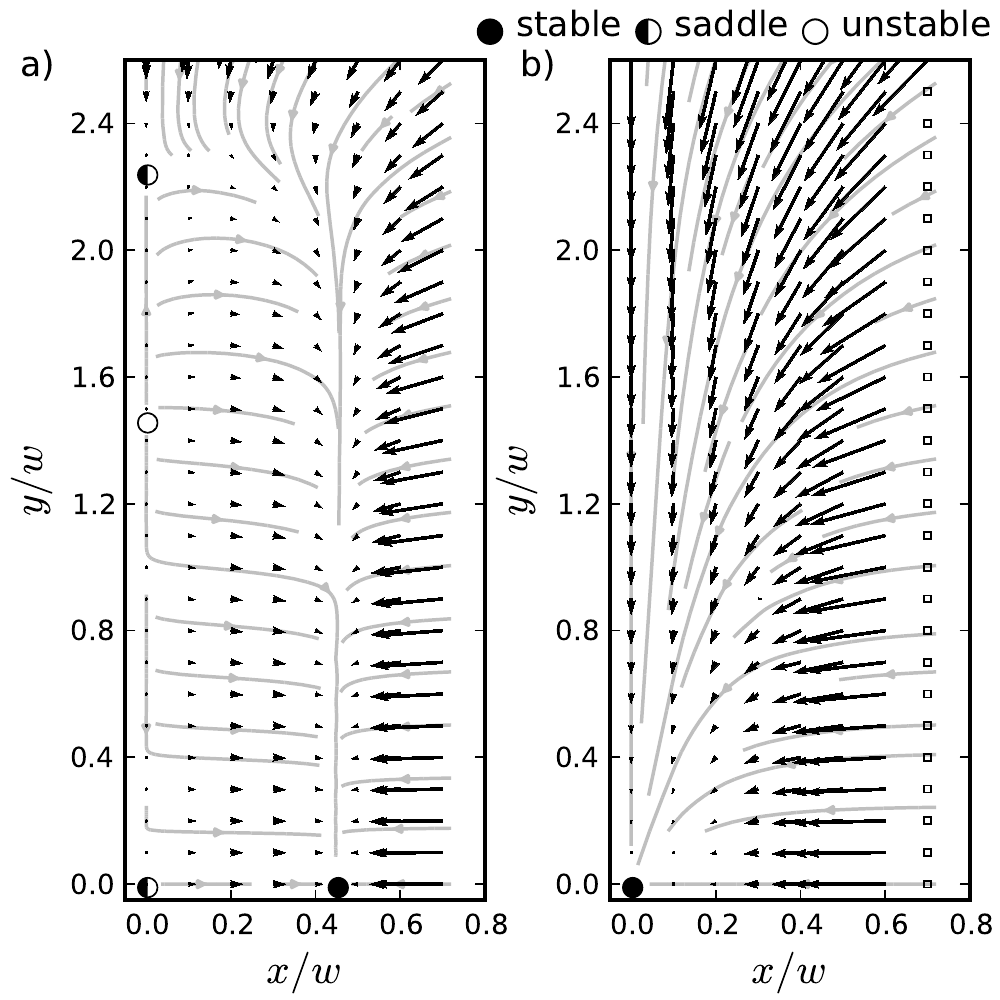}
\caption{Lift force profile for $a/w = 0.2, \mathrm{Re} = 10, w/h = 1/3$ without axial control (a) 
		 and with axial control force $f_\mathrm{ctl} = 2.5 \rho \nu^2$ (b).}
\label{fig:control-tableau}
\end{figure}

For a particle with radius $a = 0.2w$ we observe without axial control the cross sectional
force profile shown in Fig.\ \ref{fig:control-tableau}(a).
As discussed in the previous section we observe that a
particle is pushed towards the $y=0$ plane where it stays confined.
When we apply an additional axial control force of $f_\mathrm{ctl} = 2.5 \rho\nu^2$, 
the force profile changes drastically [Fig.\ \ref{fig:control-tableau}(b)]. 
In particular, the stable equilibrium position at $x/w \approx 0.46$ vanishes
and the particle is focussed to the channel center regardless of its initial position.

\begin{figure}
\includegraphics[width=\columnwidth]{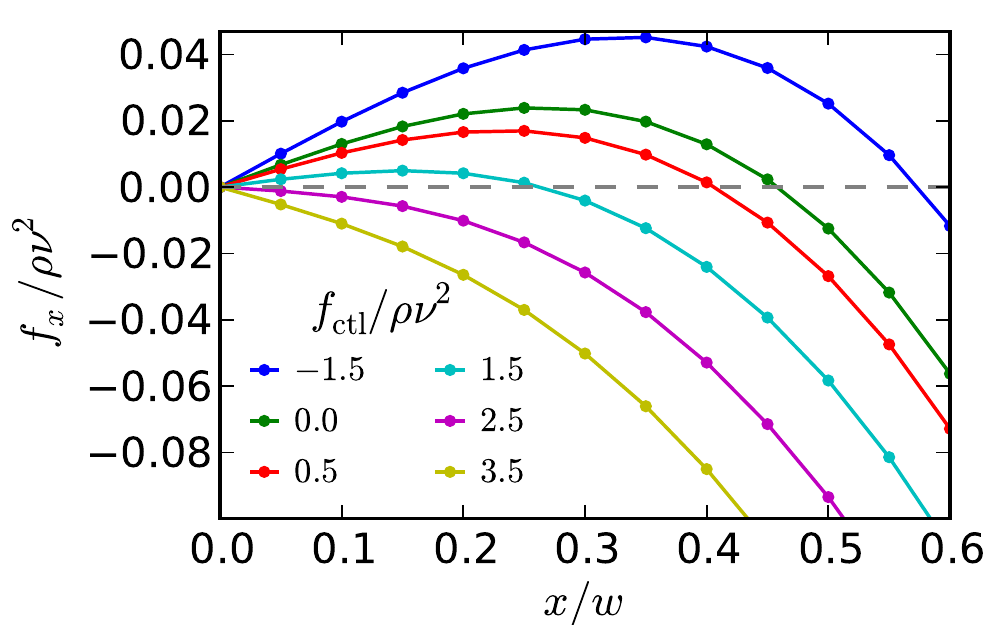}
\caption{Inertial lift force plotted versus location $x$ for different axial control forces $f_{\mathrm{ctl}}$
 for $\mathrm{Re} = 10$ and $a/w = 0.2$.}
\label{fig:control-force}
\end{figure}

We consider in the following a channel with aspect ratio $w/h = 1/3$ to ensure that
the particle is confined to the $y=0$ plane and focus on the lift force along the $x$ direction. 
In Fig.\ \ref{fig:control-force} we plot intertial lift force profiles for several axial control forces $f_{ctl}$.
For zero control force (green curve), the typical lift force profile occurs with the unstable equilibrium position
at the center and the stable position half way between channel center and wall. When we apply the axial
control force in flow direction so that the particle is sped up relative to the flow (negative $f_{\mathrm{ctl}}$), the lift force increases
and the stable equilibrium position is pushed further towards the channel wall. The stable position moves closer to the center, 
when we slow down the particle with a positive control force $f_{\mathrm{ctl}}$ acting aginst the flow. The additional Saffman force
decreases the lift force. Ultimately both fixpoints merge and the particle position at the center is stabilized when the
inertial force profile becomes completely negative.
We also studied the change in the lift force for fixed position $x$ and found that it is nearly
linear in the applied control forces. From Fig.\ \ref{fig:control-force} we already sense that the variation
of the inertia lift force with the axial control force is strongest close to the wall.


\begin{figure}
\includegraphics[width=\columnwidth]{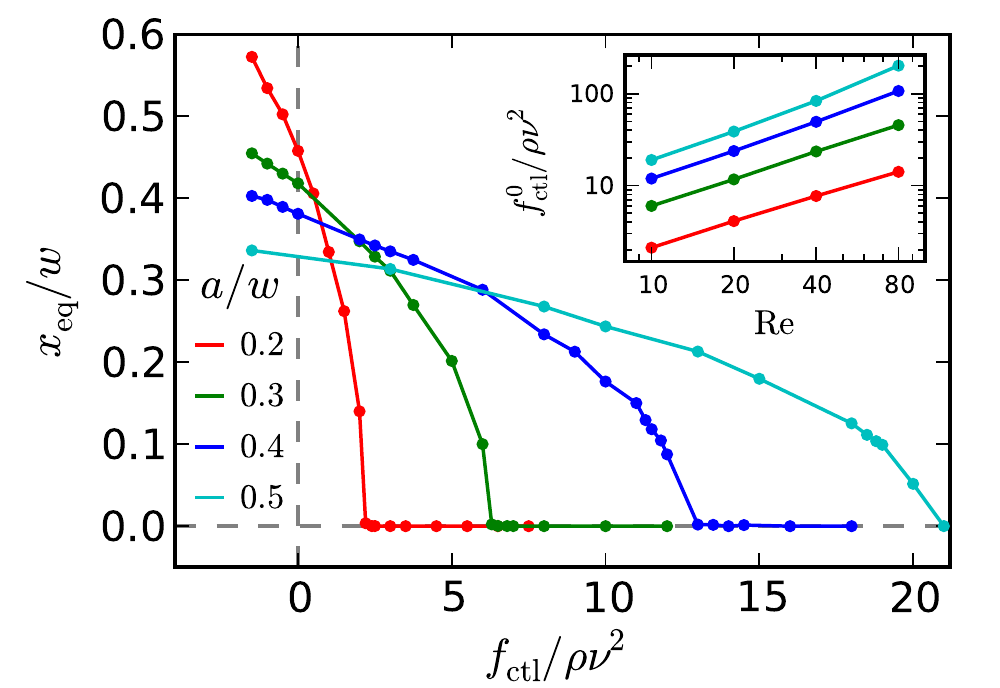}
\caption{ 
Equilibrium position $x_{\mathrm{eq}}$ plotted versus axial control force $f_{\mathrm{ctl}}$ for different
particle radii $a/w$.
The inset shows the mimimum control force $f_{\mathrm{ctl}}^{0}$ needed to focus a particle to the channel center
plotted versus Re for different particle sizes. 
}
\label{fig:control-summary}
\end{figure}

In Fig.\ \ref{fig:control-summary} we plot the stable equilibrium position $x_{\mathrm{eq}}$ versus $f_{\mathrm{ctl}}$ 
for different particle sizes.
Starting from its uncontrolled value, the equilibrium position continuously shifts to zero
as the control force increases. 
A negative control forces moves $x_{\mathrm{eq}}$ towards the wall.
In general, we observe that larger particles require larger control forces in order to move the equilibrium position.
To analyze this effect further, we investigate the minimum control force $f_{\mathrm{ctl}}^{0}$ needed to steer 
a particle towards the channel center for different Reynolds numbers and particle sizes in the inset of Fig.\ \ref{fig:control-summary}.
We observe that $f_{\mathrm{ctl}}^{0}$ strongly increases both with particle size and Reynolds number. 
When fitted to a power law, we obtain $f_\mathrm{ctl}^{0} \propto \mathrm{Re}^{1.02} (a/w)^{2.60}$.
This indicates that we can easily exploit an axial control force for particle sorting by size.
For example, consider two particle types with sizes $a/w = 0.2$ and $a/w = 0.3$.
While the small particle is well focussed to the channel center 
with a  control force $f_\mathrm{ctl} = 3\rho\nu^2$, the larger particle 
only changes its equilibrium position by about 10\,\%.

\section{Feedback control}
\label{sec:feedback}

We have seen in the previous section, already a constant axial control force 
allows to manipulate and sort particles of different types. In the following we demonstrate 
how a simple feedback scheme adds additional control to
the system and, in particluar, increases particle throughput. 
We present results, where we simulated particle motion using the Langevin 
dynamics described in Sec.~\ref{sec:langevin-equation}. 

\begin{figure}
\includegraphics[width=\columnwidth]{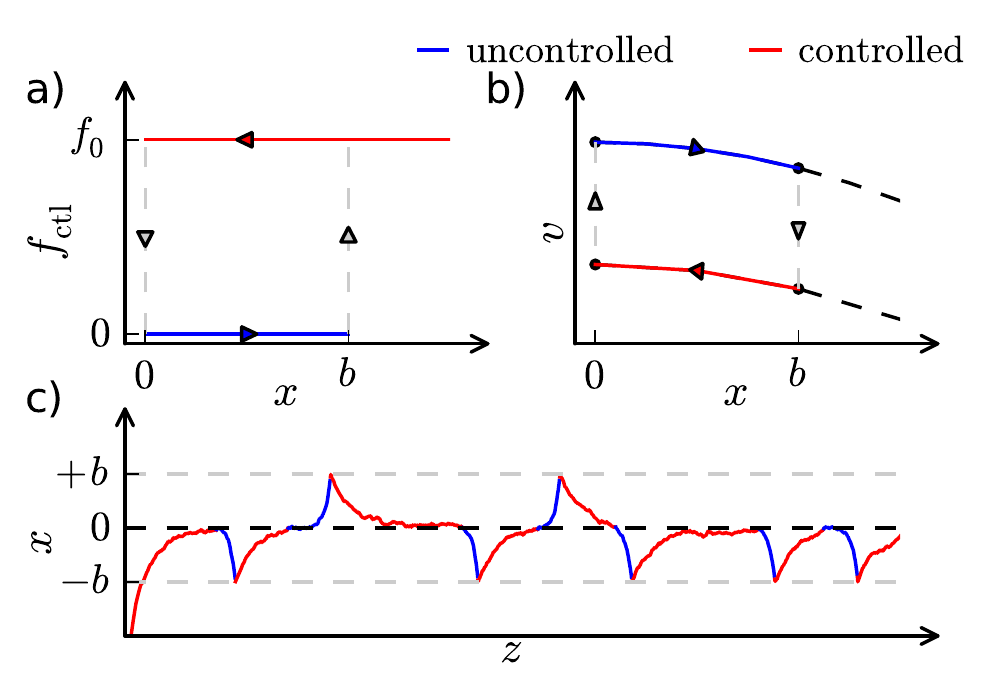}
\caption{
a) Schematic of the feedback control scheme.
The axial control force $f_0$ is switched on, when the particle leaves the target interval $[-b, b|$
and switched off once the particle reaches the centerline at $x = 0$.
(b) Particle velocity in the target interval without and under control.
(c) Example of a particle trajectory. The residence time on the centerline is determined by thermal fluctuations.
}
\label{fig:feedback-scheme}
\end{figure} 


We use a hysteretic control feedback scheme, which switches from no control to constant control depending on the 
lateral particle position with the goal to keep the particle close to the channel center.
In concrete, we choose a target interval $ [-b, b]$ for the $x$ position of the particle. 
We switch the axial control force to a constant value $f_0$, when the particle is outside the target interval. 
The modified lift force profile drives the particle back to the channel center and we switch off the control 
force until the particle leaves the target interval again.
We sketch the resulting hysteretic control cycle for $f_\mathrm{ctl}$ in Fig.~\ref{fig:feedback-scheme}(a),
 which either acts in the positive or negative $x$ direction.
The applied control force not only changes the lift force profile, but also the 
particle velocity along the channel axis,  as we  demonstrate in Fig.\ \ref{fig:feedback-scheme}(b).
When the control is active, the particle is slowed down compared to the uncontrolled motion.   
Figure\ \ref{fig:feedback-scheme}(c) shows an example of a particle trajectory under the feedback
scheme. 
The particle starts outside the target interval $[-b, b]$ and the lift force modifed by the axial control force pulls
it towards the channel centerline.
As the particle reaches the centerline, control is switched off and the particle is free to evolve. 
Since the centerline is an unstable fixpoint, the particle can leave the target interval and control is activated again.

\begin{figure}
\includegraphics[width=\columnwidth]{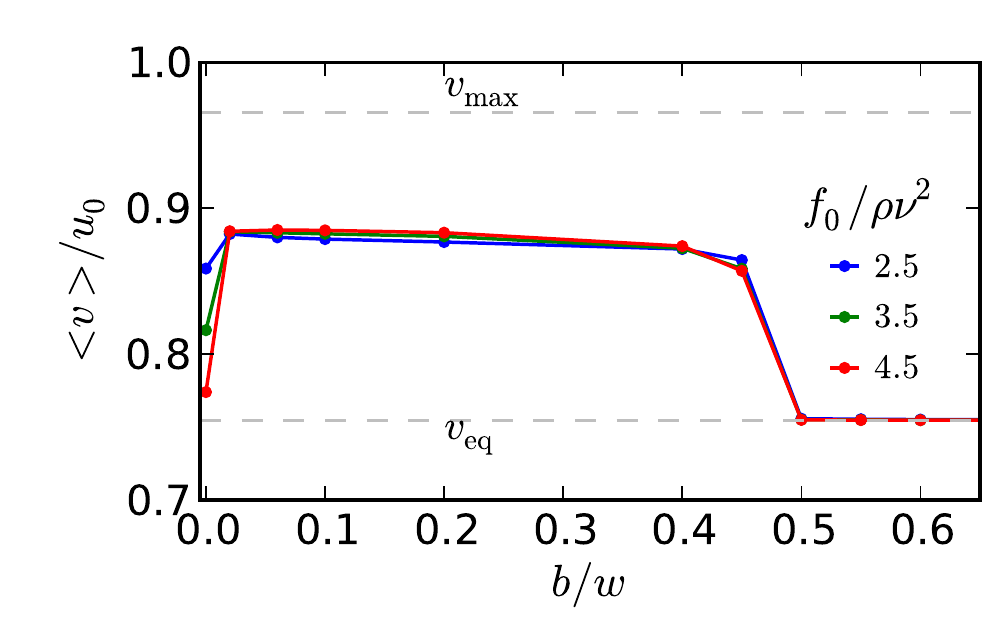}
\caption{
Mean particle velocity $\langle v \rangle$ in units of the maximum Poiseuille flow speed $u_0$
plotted versus target width $b$. Different control forces $f_0$ within the hysteretic feedback control
scheme are used. $v_{\mathrm{max}}$ is the particle speed on the centerline and $v_{\mathrm{eq}}$ 
the velocity at the equilibrium position, when no control acts.
}
\label{fig:compare-schemes}
\end{figure} 

Performing feedback control instead of using a permanently applied control force has the advantage that the
effect of control is reduced. In particular, feedback control gives an improved particle throughput while keeping the particle close 
to the centerline. In Fig.~\ref{fig:compare-schemes} we show the mean particle speed (in units of the maximal flow speed $u_0$)
as a function of the target width $b$ for several control forces $f_0$. For uncontrolled motion, $v_{\mathrm{max}}$ is the particle 
speed on the centerline and $v_{\mathrm{eq}}$ the velocity at the equilibrium position. When the target interval contains
the equilibrium position at $x=0.46$, feedback control is not active. The particle stays at the equilibrium position and 
moves with $v_{\mathrm{eq}}$. However, for smaller target widths feedback control sets in. We observe a mean particle 
velocity which is almost independent of the target width and the applied control force $f_0$. Only at $b=0$, which means 
a permanently applied control force where the particle always stays in the center, does the mean velocity decrease. 
So the particle throughput is largest when feedback control is active.

\section{Conclusion}
\label{sec.concl}

Inertial microfluidics has proven to be particularly useful for applications such as particle
steering and sorting which are important tasks in biomedical applications. In this paper we provided
further theoretical insights into inertial microfluidics using lattice Boltzmann simulations. 
We put special emphasis on controlling particle motion either by designing the channel geometry or 
by applying an additional control force.

We first investigated the equilibrium positions in square and rectangular channels using lift force
profiles and categorized them into stable, saddle, and unstable fixpoints. In square channels the stable
fixpoints either sit on the diagonals or the main axis of the cross section. This depends on particle size 
and Reynolds number and thereby offers the possibility to sort particles of different size. For rectangular
channels we illustrated bifurcation szenarios for fixpoints situated on the long main axis. In particular, we showed 
that for sufficiently elongated channel cross section particles are pushed into a plane. Their
dynamics becomes two-dimensional which simplifies the monitoring and thereby control of particle
motion in experiments.

We then demonstrated how an additional axial control force allows to tune the stable equilibrium position,
which moves towards the center with increasing force. Ultimately the stable position stays on the centerline 
when the force exceeds a threshold value which we identified for different particle sizes and Reynolds 
numbers. The strong dependence on these parameters allows to separate particles by size.
Finally, we proposed a hysteretic feedback scheme using the axial control force to enhance particle 
throughput compared to the case when the control force is constantly applied.

We plan to extend our work to investigate the collective colloidal dynamics induced by hydrodynamic
interactions between particles. Here additional axial order such as particle trains develop 
\cite{Lee2010,Humphry2010Axial}. Furthermore, it will be challenging to generalize our control methods
to particle suspensions. The theoretical insights developed in this paper and future work on
the collective dynamics will help to generate novel ideas for devices in biomedical applications based on
inertial microfluidics.

\section{Acknowledgments}
We acknowledge support by the Deutsche Forschungsgemeinschaft in the framework of the 
collaborative research center SFB 910.

\footnotesize{
\bibliography{paper}

\providecommand*{\mcitethebibliography}{\thebibliography}
\csname @ifundefined\endcsname{endmcitethebibliography}
{\let\endmcitethebibliography\endthebibliography}{}
\begin{mcitethebibliography}{41}
\providecommand*{\natexlab}[1]{#1}
\providecommand*{\mciteSetBstSublistMode}[1]{}
\providecommand*{\mciteSetBstMaxWidthForm}[2]{}
\providecommand*{\mciteBstWouldAddEndPuncttrue}
  {\def\EndOfBibitem{\unskip.}}
\providecommand*{\mciteBstWouldAddEndPunctfalse}
  {\let\EndOfBibitem\relax}
\providecommand*{\mciteSetBstMidEndSepPunct}[3]{}
\providecommand*{\mciteSetBstSublistLabelBeginEnd}[3]{}
\providecommand*{\EndOfBibitem}{}
\mciteSetBstSublistMode{f}
\mciteSetBstMaxWidthForm{subitem}
{(\emph{\alph{mcitesubitemcount}})}
\mciteSetBstSublistLabelBeginEnd{\mcitemaxwidthsubitemform\space}
{\relax}{\relax}

\bibitem[Hur \emph{et~al.}(2011)Hur, Mach, and Di~Carlo]{Hur2011}
S.~C. Hur, A.~J. Mach and D.~Di~Carlo, \emph{Biomicrofluidics}, 2011,
  \textbf{5}, 022206\relax
\mciteBstWouldAddEndPuncttrue
\mciteSetBstMidEndSepPunct{\mcitedefaultmidpunct}
{\mcitedefaultendpunct}{\mcitedefaultseppunct}\relax
\EndOfBibitem
\bibitem[Mach and Di~Carlo(2010)]{Mach2010}
A.~J. Mach and D.~Di~Carlo, \emph{Biotechnol. Bioeng.}, 2010, \textbf{107},
  302--311\relax
\mciteBstWouldAddEndPuncttrue
\mciteSetBstMidEndSepPunct{\mcitedefaultmidpunct}
{\mcitedefaultendpunct}{\mcitedefaultseppunct}\relax
\EndOfBibitem
\bibitem[Guan \emph{et~al.}(2013)Guan, Wu, Bhagat, Li, Chen, Chao, Ong, and
  Han]{Guan2013Spiral}
G.~Guan, L.~Wu, A.~A. Bhagat, Z.~Li, P.~C. Chen, S.~Chao, C.~J. Ong and J.~Han,
  \emph{Sci. Rep.}, 2013, \textbf{3}, 1475\relax
\mciteBstWouldAddEndPuncttrue
\mciteSetBstMidEndSepPunct{\mcitedefaultmidpunct}
{\mcitedefaultendpunct}{\mcitedefaultseppunct}\relax
\EndOfBibitem
\bibitem[Dudani \emph{et~al.}(2013)Dudani, Gossett, Henry, and
  Di~Carlo]{Dudani2013Pinched}
J.~S. Dudani, D.~R. Gossett, T.~Henry and D.~Di~Carlo, \emph{Lab Chip}, 2013,
  \textbf{13}, 3728--3734\relax
\mciteBstWouldAddEndPuncttrue
\mciteSetBstMidEndSepPunct{\mcitedefaultmidpunct}
{\mcitedefaultendpunct}{\mcitedefaultseppunct}\relax
\EndOfBibitem
\bibitem[Segr\'{e} and Silberberg(1961)]{Segre1961}
G.~Segr\'{e} and A.~Silberberg, \emph{Nature}, 1961, \textbf{189}, 209 --
  210\relax
\mciteBstWouldAddEndPuncttrue
\mciteSetBstMidEndSepPunct{\mcitedefaultmidpunct}
{\mcitedefaultendpunct}{\mcitedefaultseppunct}\relax
\EndOfBibitem
\bibitem[Di~Carlo \emph{et~al.}(2009)Di~Carlo, Edd, Humphry, Stone, and
  Toner]{DiCarlo2009}
D.~Di~Carlo, J.~F. Edd, K.~J. Humphry, H.~A. Stone and M.~Toner, \emph{Phys.
  Rev. Lett.}, 2009, \textbf{102}, 094503\relax
\mciteBstWouldAddEndPuncttrue
\mciteSetBstMidEndSepPunct{\mcitedefaultmidpunct}
{\mcitedefaultendpunct}{\mcitedefaultseppunct}\relax
\EndOfBibitem
\bibitem[Di~Carlo(2009)]{DiCarlo2009Inertial}
D.~Di~Carlo, \emph{Lab Chip}, 2009, \textbf{9}, 3038 -- 3046\relax
\mciteBstWouldAddEndPuncttrue
\mciteSetBstMidEndSepPunct{\mcitedefaultmidpunct}
{\mcitedefaultendpunct}{\mcitedefaultseppunct}\relax
\EndOfBibitem
\bibitem[Kataoka and Inamuro(2011)]{Kataoka2011Numerical}
Y.~Kataoka and T.~Inamuro, \emph{Phil. Trans. R. Soc. A}, 2011, \textbf{369},
  2528--2536\relax
\mciteBstWouldAddEndPuncttrue
\mciteSetBstMidEndSepPunct{\mcitedefaultmidpunct}
{\mcitedefaultendpunct}{\mcitedefaultseppunct}\relax
\EndOfBibitem
\bibitem[Chun and Ladd(2006)]{Chun2006}
B.~Chun and A.~J.~C. Ladd, \emph{Phys. Fluids}, 2006, \textbf{18}, 031704\relax
\mciteBstWouldAddEndPuncttrue
\mciteSetBstMidEndSepPunct{\mcitedefaultmidpunct}
{\mcitedefaultendpunct}{\mcitedefaultseppunct}\relax
\EndOfBibitem
\bibitem[Zhou and Papautsky(2013)]{Zhou2013Fundamentals}
J.~Zhou and I.~Papautsky, \emph{Lab Chip}, 2013,  1121 -- 1132\relax
\mciteBstWouldAddEndPuncttrue
\mciteSetBstMidEndSepPunct{\mcitedefaultmidpunct}
{\mcitedefaultendpunct}{\mcitedefaultseppunct}\relax
\EndOfBibitem
\bibitem[Matas \emph{et~al.}(2004)Matas, Morris, and Guazzelli]{Matas2004}
J.~P. Matas, J.~F. Morris and E.~Guazzelli, \emph{Oil Gas Sci. Technol.}, 2004,
  \textbf{59}, 59--70\relax
\mciteBstWouldAddEndPuncttrue
\mciteSetBstMidEndSepPunct{\mcitedefaultmidpunct}
{\mcitedefaultendpunct}{\mcitedefaultseppunct}\relax
\EndOfBibitem
\bibitem[Bruus(2007)]{BookBruus}
H.~Bruus, \emph{Theoretical microfluidics}, Oxford University Press, 2007\relax
\mciteBstWouldAddEndPuncttrue
\mciteSetBstMidEndSepPunct{\mcitedefaultmidpunct}
{\mcitedefaultendpunct}{\mcitedefaultseppunct}\relax
\EndOfBibitem
\bibitem[Asmolov(1999)]{Asmolov1999}
E.~S. Asmolov, \emph{J. Fluid Mech.}, 1999, \textbf{381}, 63--87\relax
\mciteBstWouldAddEndPuncttrue
\mciteSetBstMidEndSepPunct{\mcitedefaultmidpunct}
{\mcitedefaultendpunct}{\mcitedefaultseppunct}\relax
\EndOfBibitem
\bibitem[Ho and Leal(1974)]{Ho1974}
B.~P. Ho and L.~G. Leal, \emph{J. Fluid Mech.}, 1974, \textbf{65},
  365--400\relax
\mciteBstWouldAddEndPuncttrue
\mciteSetBstMidEndSepPunct{\mcitedefaultmidpunct}
{\mcitedefaultendpunct}{\mcitedefaultseppunct}\relax
\EndOfBibitem
\bibitem[Prohm \emph{et~al.}(2012)Prohm, Gierlak, and Stark]{Prohm2012}
C.~Prohm, M.~Gierlak and H.~Stark, \emph{Eur. Phys. J. E}, 2012, \textbf{35},
  1--10\relax
\mciteBstWouldAddEndPuncttrue
\mciteSetBstMidEndSepPunct{\mcitedefaultmidpunct}
{\mcitedefaultendpunct}{\mcitedefaultseppunct}\relax
\EndOfBibitem
\bibitem[MacDonald \emph{et~al.}(2003)MacDonald, Spalding, and
  Dholakia]{MacDonald2003}
M.~MacDonald, G.~Spalding and K.~Dholakia, \emph{Nature}, 2003, \textbf{426},
  421--424\relax
\mciteBstWouldAddEndPuncttrue
\mciteSetBstMidEndSepPunct{\mcitedefaultmidpunct}
{\mcitedefaultendpunct}{\mcitedefaultseppunct}\relax
\EndOfBibitem
\bibitem[Prohm \emph{et~al.}(2013)Prohm, Tröltzsch, and Stark]{Prohm2013}
C.~Prohm, F.~Tröltzsch and H.~Stark, \emph{Eur. Phys. J. E}, 2013,
  \textbf{36}, 1--13\relax
\mciteBstWouldAddEndPuncttrue
\mciteSetBstMidEndSepPunct{\mcitedefaultmidpunct}
{\mcitedefaultendpunct}{\mcitedefaultseppunct}\relax
\EndOfBibitem
\bibitem[Kim and Yoo(2009)]{Kim2009}
W.~Y. Kim and J.~Y. Yoo, \emph{Lab Chip}, 2009, \textbf{9}, 1043 -- 1045\relax
\mciteBstWouldAddEndPuncttrue
\mciteSetBstMidEndSepPunct{\mcitedefaultmidpunct}
{\mcitedefaultendpunct}{\mcitedefaultseppunct}\relax
\EndOfBibitem
\bibitem[Saffman(1965)]{Saffman1965}
P.~G. Saffman, \emph{J. Fluid Mech.}, 1965, \textbf{22}, 384 -- 400\relax
\mciteBstWouldAddEndPuncttrue
\mciteSetBstMidEndSepPunct{\mcitedefaultmidpunct}
{\mcitedefaultendpunct}{\mcitedefaultseppunct}\relax
\EndOfBibitem
\bibitem[Astr{\"o}m and Murray(2010)]{BookAstrom2010Feedback}
K.~Astr{\"o}m and R.~Murray, \emph{Feedback Systems: An Introduction for
  Scientists and Engineers}, Princeton University Press, 2010\relax
\mciteBstWouldAddEndPuncttrue
\mciteSetBstMidEndSepPunct{\mcitedefaultmidpunct}
{\mcitedefaultendpunct}{\mcitedefaultseppunct}\relax
\EndOfBibitem
\bibitem[Jon{\'a}{\v{s}} and Zem{\'a}nek(2008)]{Jonas2008Light}
A.~Jon{\'a}{\v{s}} and P.~Zem{\'a}nek, \emph{Electrophoresis}, 2008,
  \textbf{29}, 4813--4851\relax
\mciteBstWouldAddEndPuncttrue
\mciteSetBstMidEndSepPunct{\mcitedefaultmidpunct}
{\mcitedefaultendpunct}{\mcitedefaultseppunct}\relax
\EndOfBibitem
\bibitem[Wuite \emph{et~al.}(2000)Wuite, Smith, Young, Keller, and
  Bustamante]{Wuite2000Single}
G.~J. Wuite, S.~B. Smith, M.~Young, D.~Keller and C.~Bustamante, \emph{Nature},
  2000, \textbf{404}, 103--106\relax
\mciteBstWouldAddEndPuncttrue
\mciteSetBstMidEndSepPunct{\mcitedefaultmidpunct}
{\mcitedefaultendpunct}{\mcitedefaultseppunct}\relax
\EndOfBibitem
\bibitem[Wang \emph{et~al.}(1997)Wang, Yin, Landick, Gelles, and
  Block]{Wang1997Stretching}
M.~D. Wang, H.~Yin, R.~Landick, J.~Gelles and S.~M. Block, \emph{Biophys. J.},
  1997, \textbf{72}, 1335--1346\relax
\mciteBstWouldAddEndPuncttrue
\mciteSetBstMidEndSepPunct{\mcitedefaultmidpunct}
{\mcitedefaultendpunct}{\mcitedefaultseppunct}\relax
\EndOfBibitem
\bibitem[Applegate~Jr \emph{et~al.}(2007)Applegate~Jr, Schafer, Amir, Squier,
  Vestad, Oakey, and Marr]{Applegate2007Optically}
R.~W. Applegate~Jr, D.~N. Schafer, W.~Amir, J.~Squier, T.~Vestad, J.~Oakey and
  D.~W. Marr, \emph{J. Opt. A - Pure Appl. Op.}, 2007, \textbf{9}, S122\relax
\mciteBstWouldAddEndPuncttrue
\mciteSetBstMidEndSepPunct{\mcitedefaultmidpunct}
{\mcitedefaultendpunct}{\mcitedefaultseppunct}\relax
\EndOfBibitem
\bibitem[Wang \emph{et~al.}(2011)Wang, Chen, Kong, Wang, Costa, Li, and
  Sun]{Wang2011Enhanced}
X.~Wang, S.~Chen, M.~Kong, Z.~Wang, K.~D. Costa, R.~A. Li and D.~Sun, \emph{Lab
  Chip}, 2011, \textbf{11}, 3656--3662\relax
\mciteBstWouldAddEndPuncttrue
\mciteSetBstMidEndSepPunct{\mcitedefaultmidpunct}
{\mcitedefaultendpunct}{\mcitedefaultseppunct}\relax
\EndOfBibitem
\bibitem[Munson \emph{et~al.}(2010)Munson, Spotts, Niemist{\"o}, Selinummi,
  Kralj, Salit, and Ozinsky]{Munson2010Image}
M.~S. Munson, J.~M. Spotts, A.~Niemist{\"o}, J.~Selinummi, J.~G. Kralj, M.~L.
  Salit and A.~Ozinsky, \emph{Lab Chip}, 2010, \textbf{10}, 2402--2410\relax
\mciteBstWouldAddEndPuncttrue
\mciteSetBstMidEndSepPunct{\mcitedefaultmidpunct}
{\mcitedefaultendpunct}{\mcitedefaultseppunct}\relax
\EndOfBibitem
\bibitem[Dünweg and Ladd(2008)]{Duenweg2008Lattice}
B.~Dünweg and A.~J. Ladd, \emph{Advances in Polymer Science}, Springer Berlin
  Heidelberg, 2008, pp. 1--78\relax
\mciteBstWouldAddEndPuncttrue
\mciteSetBstMidEndSepPunct{\mcitedefaultmidpunct}
{\mcitedefaultendpunct}{\mcitedefaultseppunct}\relax
\EndOfBibitem
\bibitem[Aidun and Clausen(2010)]{Aidun2010Lattice}
C.~K. Aidun and J.~R. Clausen, \emph{Annu. Rev. Fluid Mech.}, 2010,
  \textbf{42}, 439--472\relax
\mciteBstWouldAddEndPuncttrue
\mciteSetBstMidEndSepPunct{\mcitedefaultmidpunct}
{\mcitedefaultendpunct}{\mcitedefaultseppunct}\relax
\EndOfBibitem
\bibitem[Latt \emph{et~al.}(2008)Latt, Chopard, Malaspinas, Deville, and
  Michler]{Latt2008}
J.~Latt, B.~Chopard, O.~Malaspinas, M.~Deville and A.~Michler, \emph{Phys. Rev.
  E}, 2008, \textbf{77}, 056703\relax
\mciteBstWouldAddEndPuncttrue
\mciteSetBstMidEndSepPunct{\mcitedefaultmidpunct}
{\mcitedefaultendpunct}{\mcitedefaultseppunct}\relax
\EndOfBibitem
\bibitem[Zou and He(1997)]{Zou1997Pressure}
Q.~Zou and X.~He, \emph{Phys. Fluids}, 1997, \textbf{9}, 1591\relax
\mciteBstWouldAddEndPuncttrue
\mciteSetBstMidEndSepPunct{\mcitedefaultmidpunct}
{\mcitedefaultendpunct}{\mcitedefaultseppunct}\relax
\EndOfBibitem
\bibitem[Shan and Chen(1993)]{Shan1993}
X.~Shan and H.~Chen, \emph{Phys. Rev. E}, 1993, \textbf{47}, 1815--1819\relax
\mciteBstWouldAddEndPuncttrue
\mciteSetBstMidEndSepPunct{\mcitedefaultmidpunct}
{\mcitedefaultendpunct}{\mcitedefaultseppunct}\relax
\EndOfBibitem
\bibitem[Inamuro(2012)]{Inamuro2012Lattice}
T.~Inamuro, \emph{Fluid Dyn. Res.}, 2012, \textbf{44}, 024001\relax
\mciteBstWouldAddEndPuncttrue
\mciteSetBstMidEndSepPunct{\mcitedefaultmidpunct}
{\mcitedefaultendpunct}{\mcitedefaultseppunct}\relax
\EndOfBibitem
\bibitem[Peskin(2002)]{Peskin2002}
C.~S. Peskin, \emph{Acta numerica}, 2002, \textbf{11}, 479--517\relax
\mciteBstWouldAddEndPuncttrue
\mciteSetBstMidEndSepPunct{\mcitedefaultmidpunct}
{\mcitedefaultendpunct}{\mcitedefaultseppunct}\relax
\EndOfBibitem
\bibitem[Feng and Michaelides(2009)]{Feng2009Robust}
Z.-G. Feng and E.~E. Michaelides, \emph{Comput. Fluids}, 2009, \textbf{38},
  370--381\relax
\mciteBstWouldAddEndPuncttrue
\mciteSetBstMidEndSepPunct{\mcitedefaultmidpunct}
{\mcitedefaultendpunct}{\mcitedefaultseppunct}\relax
\EndOfBibitem
\bibitem[Pal(2013)]{PalabosWebsite}
\emph{The Palabos project}, 2013, \url{http://www.palabos.org}\relax
\mciteBstWouldAddEndPuncttrue
\mciteSetBstMidEndSepPunct{\mcitedefaultmidpunct}
{\mcitedefaultendpunct}{\mcitedefaultseppunct}\relax
\EndOfBibitem
\bibitem[Le and Zhang(2009)]{Le2009Boundary}
G.~Le and J.~Zhang, \emph{Phys. Rev. E}, 2009, \textbf{79}, 026701\relax
\mciteBstWouldAddEndPuncttrue
\mciteSetBstMidEndSepPunct{\mcitedefaultmidpunct}
{\mcitedefaultendpunct}{\mcitedefaultseppunct}\relax
\EndOfBibitem
\bibitem[Kr{\"u}ger \emph{et~al.}(2009)Kr{\"u}ger, Varnik, and
  Raabe]{Kruger2009Shear}
T.~Kr{\"u}ger, F.~Varnik and D.~Raabe, \emph{Phys. Rev. E}, 2009, \textbf{79},
  046704\relax
\mciteBstWouldAddEndPuncttrue
\mciteSetBstMidEndSepPunct{\mcitedefaultmidpunct}
{\mcitedefaultendpunct}{\mcitedefaultseppunct}\relax
\EndOfBibitem
\bibitem[Kloeden and Platen(2011)]{BookKloeden}
P.~Kloeden and E.~Platen, \emph{Numerical Solution of Stochastic Differential
  Equations}, Springer, 2011\relax
\mciteBstWouldAddEndPuncttrue
\mciteSetBstMidEndSepPunct{\mcitedefaultmidpunct}
{\mcitedefaultendpunct}{\mcitedefaultseppunct}\relax
\EndOfBibitem
\bibitem[Hur \emph{et~al.}(2010)Hur, Tse, and Di~Carlo]{Hur2010}
S.~C. Hur, H.~T.~K. Tse and D.~Di~Carlo, \emph{Lab Chip}, 2010, \textbf{10},
  274--280\relax
\mciteBstWouldAddEndPuncttrue
\mciteSetBstMidEndSepPunct{\mcitedefaultmidpunct}
{\mcitedefaultendpunct}{\mcitedefaultseppunct}\relax
\EndOfBibitem
\bibitem[Lee \emph{et~al.}(2010)Lee, Amini, Stone, and Di~Carlo]{Lee2010}
W.~Lee, H.~Amini, H.~A. Stone and D.~Di~Carlo, \emph{Proc. Natl. Acad. Sci.
  USA}, 2010, \textbf{107}, 22413--22418\relax
\mciteBstWouldAddEndPuncttrue
\mciteSetBstMidEndSepPunct{\mcitedefaultmidpunct}
{\mcitedefaultendpunct}{\mcitedefaultseppunct}\relax
\EndOfBibitem
\bibitem[Humphry \emph{et~al.}(2010)Humphry, Kulkarni, Weitz, Morris, and
  Stone]{Humphry2010Axial}
K.~J. Humphry, P.~M. Kulkarni, D.~A. Weitz, J.~F. Morris and H.~A. Stone,
  \emph{Phys. Fluid}, 2010, \textbf{22}, 081703\relax
\mciteBstWouldAddEndPuncttrue
\mciteSetBstMidEndSepPunct{\mcitedefaultmidpunct}
{\mcitedefaultendpunct}{\mcitedefaultseppunct}\relax
\EndOfBibitem
\end{mcitethebibliography}
\bibliographystyle{rsc}
}

\end{document}